\pdfoutput=1
\documentclass[11pt]{article}

\usepackage[preprint]{acl}

\usepackage{times}
\usepackage{latexsym}
\usepackage{makecell} % in the preamble
\usepackage{supertabular}

\usepackage[T1]{fontenc}
\usepackage[utf8]{inputenc}

\usepackage{microtype}

\usepackage{inconsolata}

\usepackage{graphicx}

\title{Catching Stray Balls: Football, fandom, and the impact on digital discourse}

\author{Mark J. Hill \\
  King's College London, London, United Kingdom \\
  \texttt{mark.j.hill@kcl.ac.uk} 
  }

\begin{document}
{\makeatletter\acl@finalcopytrue
  \maketitle
\begin{abstract}
This paper examines how emotional responses to football matches influence online discourse across digital spaces on Reddit. By analysing millions of posts from dozens of subreddits, it demonstrates that real-world events trigger sentiment shifts that move across communities. It shows that negative sentiment correlates with problematic language; match outcomes directly influence sentiment and posting habits; sentiment can transfer to unrelated communities; and offers insights into the content of this shifting discourse. These findings reveal how digital spaces function not as isolated environments, but as interconnected emotional ecosystems vulnerable to cross-domain contagion triggered by real-world events, contributing to our understanding of the propagation of online toxicity. While football is used as a case-study to computationally measure affective causes and movements, these patterns have implications for understanding online communities broadly.
\end{abstract}

\section{Introduction}

Football fans are often negatively presented in popular culture due to connotations of hooliganism and far-right allegiances \cite{awan_hate_2023, londonassembly_police_2023}. In response, football clubs (FCs), leagues, and fans themselves have moved to exclude those who exhibit these behaviours from matches, and instead foster tolerant atmospheres through initiatives like Kick It Out and the Rainbow Laces campaign. However, while stadiums have become more inclusive, toxicity remains online  \cite{murray_uks_2021, kassam_four_2024}. 

It is essential to emphasise that problematic supporters --- both online and offline --- represent a small minority of fans, most of whom engage positively with the sport and its communities \cite{miranda_hate_2024}. Nevertheless, the visibility and impact of this minority warrants attention. Additionally, even among well-intentioned supporters, the tribal nature of sports can occasionally lead to conflict as personal identities become entwined with team allegiances and historic rivalries \cite{sandvoss_game_2004, porat_football_2010, parry_game_2012, cleland_collective_2018}. This complex landscape, with multiple fan identities interacting, has prompted growing scholarly interest in online football communities \cite{rowe_cultures_2010, nuttall_online_2018, Woods_2022, kaden_i_2023}. However, much of this research treats online spaces as isolated environments, failing to capture how football discourse can emerge and move across digital contexts.

This paper takes a broader view by examining the affectional dynamics of football supporter communities and their movements across online spaces. Specifically, by analysing millions of posts on Reddit it, first, shows that negative sentiment correlates with problematic language; second, match outcomes influence posting habits and sentiment; third,  sentiment can transfer to unrelated communities; and fourth, offers insights into the content of these discourses. In doing this it demonstrates a direct relationship between real-world events and online discourse, offering new insights into the propagation and dynamics of online content. 

\section{Related Work}

In addition to the investigations into football fandom referenced above, the affective character of football on fans is well established both offline \cite{kerr_emotional_2005} and online \cite{wang_making_2023}, with even clubs leveraging social media to foster emotional attachment \cite{marques_sports_2018}. While the dominant emotion identified in studies of online fans is negativity, which has been shown to lead to disengagement \cite{kaden_i_2023} and hostility \cite{fenton_womens_2024}, negativity is not uniformly corrosive --- fans of underperforming teams may embrace suffering as part of loyalty \cite{newson_united_2021}. That is, negative emotions in football can be both alienating and identity-forming, making it a tricky concept to define from the fan's perspective. Nonetheless, it is common within football communities, and this research looks at how these emotions may impact digital communities.

It has been shown that users encountering hostile online discourse experience negative psychological consequences \cite{braghieri_social_2022}, while disengagement from toxic environments improves well-being \cite{allcott_welfare_2020}. However, just as negativity from a fan's perspective is difficult to conceptualise, so too is framing online negativity which takes various, and often poorly defined, forms, including hate speech, trolling, and incivility \cite{antoci_civility_2016, anderson_social_2017, matamoros-fernandez_racism_2021}. In this context "toxicity" has emerged as an umbrella term. Hanscom et al. define toxicity as "interactions directed at an entity designed to be inflammatory" \citeyearpar{hanscom_toxicity_2024}, with Recuero highlighting how it is enabled by platform affordances \citeyearpar{recuero_platformization_2024}. The latter point is of particular interest as we examine how one affordance --- inter-subreddit mobility --- facilitates emotional spillover across online spaces and influences discourse \cite{papacharissi_affective_2014}.

Social media's networked nature, combined with platform design choices, amplify emotional content \cite{milli_engagement_2025, kramer_experimental_2014}. Studies have shown that those exposed to emotionally charged social media (both positive and negative) are themselves more likely to express similar sentiments online \cite{ferrara_measuring_2015, Brady_2017}. Additionally, negative content has been shown to spread more frequently and faster than positive content \cite{tsugawa_negative_2015}. This spread has been framed in various ways as "emotional contagion" \cite{goldenberg_digital_2020}, risking feedback loops that encourage and reinforce incivility \cite{wulczyn_ex_2017}. It is the emergence of, and exposure to, negativity, that this paper investigates.

Computational detection of negative and harmful discourse is extensively studied \cite{schmidt_survey_2017, saleem_web_2017, ayo_machine_2020, jahan_systematic_2023, torregrosa_survey_2023}. However, as a subject there are significant challenges, including language ambiguity, contextual dependencies, and data biases \cite{davidson_automated_2017, sap_risk_2019, pavlopoulos_toxicity_2020,vidgen_directions_2020}.\footnote{Despite limitations, tools such as Google Jigsaw's Perspective AI are being deployed \cite{lees_new_2022}.} Within this complex computational and conceptual landscape, this approach foregrounds the multifaceted nature through which emotions emerge and are expressed online. That is, rather than proposing a method to identify harmful discourse, we aim to track it from a potential source.

\section{Data and Methods}

Our analysis examines 62,384,329 Reddit posts taken between July 2008 and August 2024 from 41 football club subreddits.\footnote{Data processing and analysis scripts are available at https://github.com/markjhill/2025-catching-strays.} Community sizes vary considerably, with a mean 1,521,569 and median 63,606 posts (Appendix \ref{sec:appendix_subreddits}). Posts were aligned with historic match results (final scores) using kick-off times plus 120-minutes (allowing for half-time breaks and added time). This covers 20,764 unique games from the top four leagues in English football, both domestic cups, the three European cup competitions, and the Community Shield (Table~\ref{tab:match_posts}).

\begin{table}
  \centering
  \resizebox{\columnwidth}{!}{
    \begin{tabular}{lrrrr}
      \hline
      \textbf{Match Result} & \textbf{Posts} & \textbf{\%} \\
      \hline
      Wins & 6,477,964 & 49.6\\
      Draws & 2,690,511 & 20.6 \\
      Losses & 3,902,686 & 29.9 \\
      Total & 13,071,161 & 100 \\
      \hline
    \end{tabular}
  }
  \caption{Match-aligned posts overview.}
  \label{tab:match_posts}
\end{table}

Posts were analysed using TweetNLP, a RoBERTa-based sentiment detection model \cite{camacho-collados_tweetnlp_2022}. Outputs (categorical probability scores) were normalised into a -1 to +1 scale to aid comparisons across contexts.\footnote{Using $S = \frac{p - n}{p + neu + n}$ where $S$ is the sentiment index, $p$ is the proportion of positive, $neu$ is the proportion of neutral, and $n$ is the proportion of negative content.} 

For cross-community analyses an additional dataset of 1,151,726 posts was constructed, made up of pairs of posts by the same user. Each pair includes one post from a club subreddit in which they are a top-1,000 poster in, and a post from a non-club subreddit made within 10-minutes of the first. These narrow windows allow us to isolate and measure sentiment transfer effects (Table~\ref{tab:sentiment_analysis}). 

\begin{table}
  \centering
  \resizebox{\columnwidth}{!}{
    \begin{tabular}{lrrrr}
      \hline
      \textbf{Post Type} & \textbf{Negative} & \textbf{Neutral} & \textbf{Positive} \\
      \hline
      FC Corpus & 36.0\% & 42.3\% & 21.7\% \\
      Paired Corpus & 37.8\% & 40.7\% & 21.5\% \\
      \;Paired FC & 38.7\% & 38.5\% & 22.8\% \\
      \;Paired non-FC & 36.9\% & 43.0\% & 20.2\% \\
      \hline
    \end{tabular}
  }
  \caption{Sentiment distribution (percentages) across subreddit post sources.}
  \label{tab:sentiment_analysis}
\end{table}

Three subsets of potentially problematic posts were constructed representing hate speech (281,110 posts), obscene language (9,823,428), and toxicity (586,222) (Table~\ref{tab:hate_datasets}). Posts were identified using two lexical datasets --- a hate speech lexicon \cite{davidson_automated_2017} and the LDNOOBW \citeyearpar{LDNOOBW2025} dataset --- and ToxicityModel \cite{nicholas22aira}, a RoBERTa-based toxicity detector.\footnote{Only 10\% of our corpus was tested with ToxicityModel due to limited computational resources.}

\begin{table}
  \centering
  \resizebox{\columnwidth}{!}{
    \begin{tabular}{lrrrr}
      \hline
      \textbf{Dataset} & \textbf{Mean} & \textbf{Median} & \textbf{SD} \\
      \hline
      FC Corpus &  -0.103& -0.108& 0.577\\
      Hate Speech &  -0.556 & -0.779 & 0.496 \\
      Obscene Language &  -0.371 & -0.613 & 0.584 \\
      Toxic & -0.523 & -0.788 & 0.535 \\
      \hline
    \end{tabular}
  }
  \caption{Mean, median and standard deviation of sentiment scores across FC Corpus and three potentially problematic post corpora.}
  \label{tab:hate_datasets}
\end{table}

With this data the paper employs a multi-layered analytical framework to systematically investigate the relationships between real-world football events, online sentiment patterns, and cross-community discourse. The approach progresses through four analytical phases: first, establishing the correlation between negative sentiment and problematic language. Second, measuring relationships between match outcomes and sentiment within FC communities. Third, identifying and correlating post sentiment across unrelated online spaces. Fourth, quantify linguistic features through lexical matching and machine learning to provide insights into the content of the identified discourses. Through this methodical progression, we establish not only statistical relationships but also temporal precedence, providing insights into the potential causal mechanisms underlying emotional spillover in digital spaces.

\section{Sentiment and Harmful Discourse}

The relationship between negative sentiment and the potentially problematic posts was assessed by comparing each datasets' overall sentiment with the FC Corpus. Low sentiment scores correlate with problematic posts (Figure \ref{fig:density_plots}).

\begin{figure}[t]
    \includegraphics[width=\columnwidth]{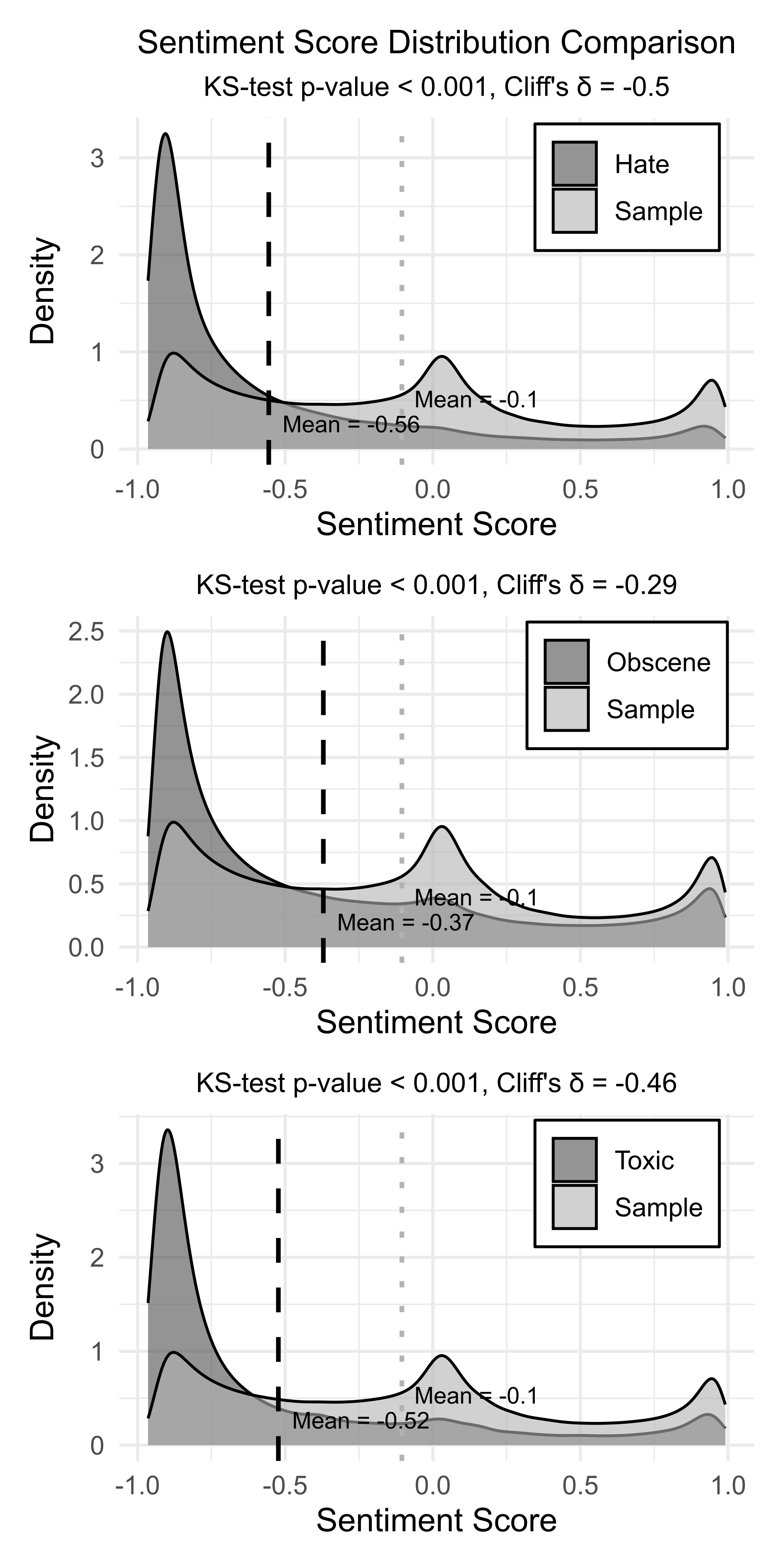}
    \caption{Sentiment score distributions comparing FC corpus and datasets of problematic content.}
    \label{fig:density_plots}
\end{figure}

Additionally, density plots show distributions between problematic posts and the FC Corpus notably differ: all three are negatively skewed and strongly unimodal, while the main corpus is balanced and multimodal. The effect sizes are strong for the hate speech and toxic datasets (Cliff's $\delta$ = -0.5, -0.47) and moderate for the obscene language dataset ($\delta$ = -0.3). The latter is visible in obscenity aligning slightly more closely with the FC Corpus' positive sentiment, indicating --- as one may expect --- that obscene language is not universally negative. However, in all three cases, one is more likely to find negative posts in the problematic datasets than the FC Corpus, suggesting sentiment may serve as a supplementary signal when detecting potentially problematic content. 

\section{Football Results as a Causal Pathway}

To make meaningful claims about the movement of sentiment across digital spaces, one must investigate the sources of sentiment. Without this, the direction of emotional contagion is difficult to assess. Football provides an ideal case-study, as it offers clear time-stamped real-world events that can be linked to subsequent posting behaviours in related online spaces. This section, therefore, assess relationships between match outcomes and sentiment patterns using posts in FC subreddits.

\subsection{Match Results and Shifts in Sentiment}

Figure \ref{fig:box-plot-change} provides evidence for shifts in sentiment within the 48-hour period around match kick-off, categorised by match result. While all three subsets have similar median sentiment before kick-off, levels after are distinct. Following losses and draws sentiment declines (-0.352, -0.225). Interestingly, wins show only a slight increase in median sentiment (0.035). 

\begin{figure}[t]
    \centering
    \includegraphics[width=\columnwidth]{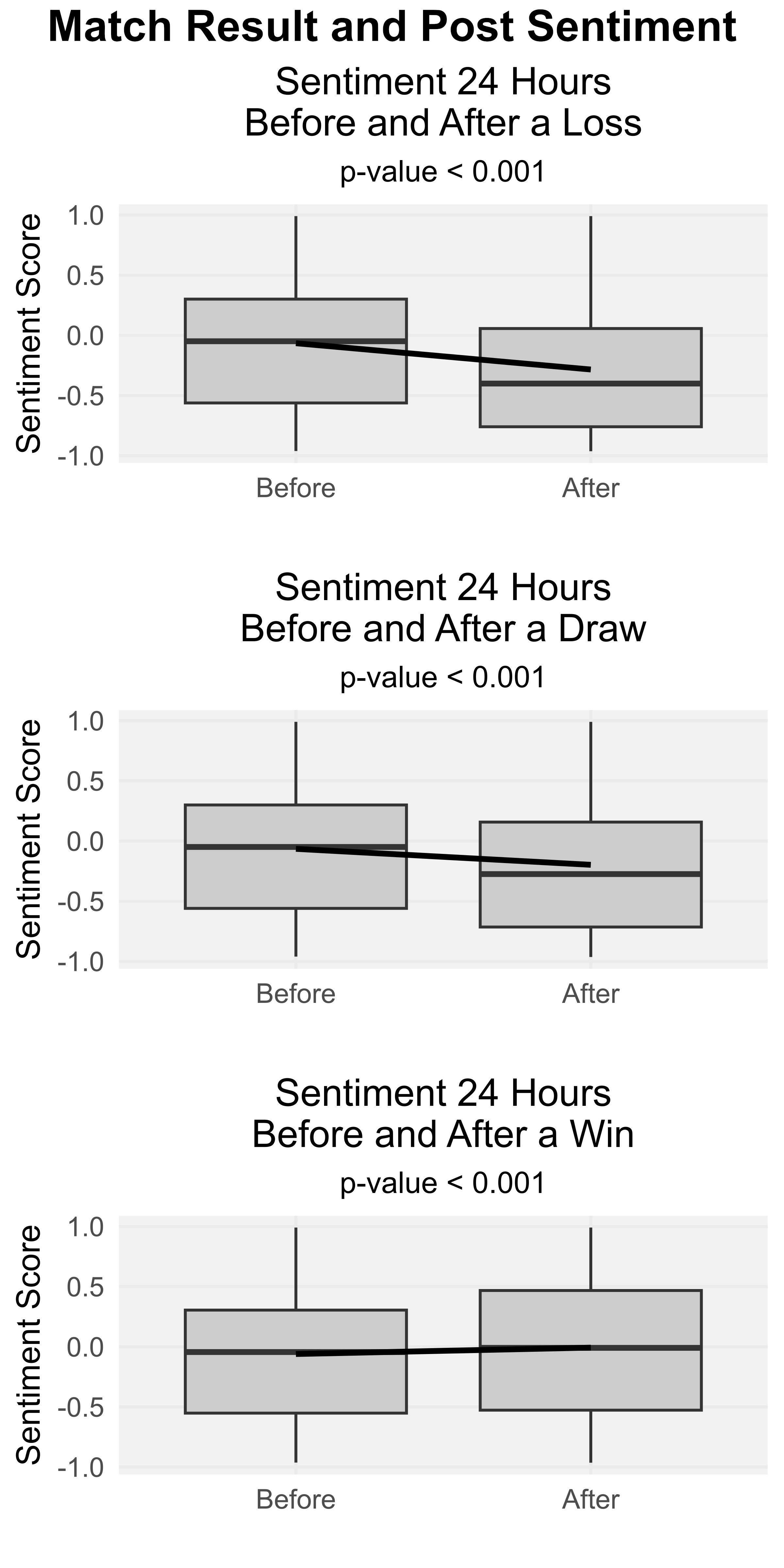}
    \caption{Change in poster sentiment over 48-hour period around kick-off (FC Corpus).}
    \label{fig:box-plot-change}
\end{figure}

\begin{figure}
    \centering
    \includegraphics[width=1\linewidth]{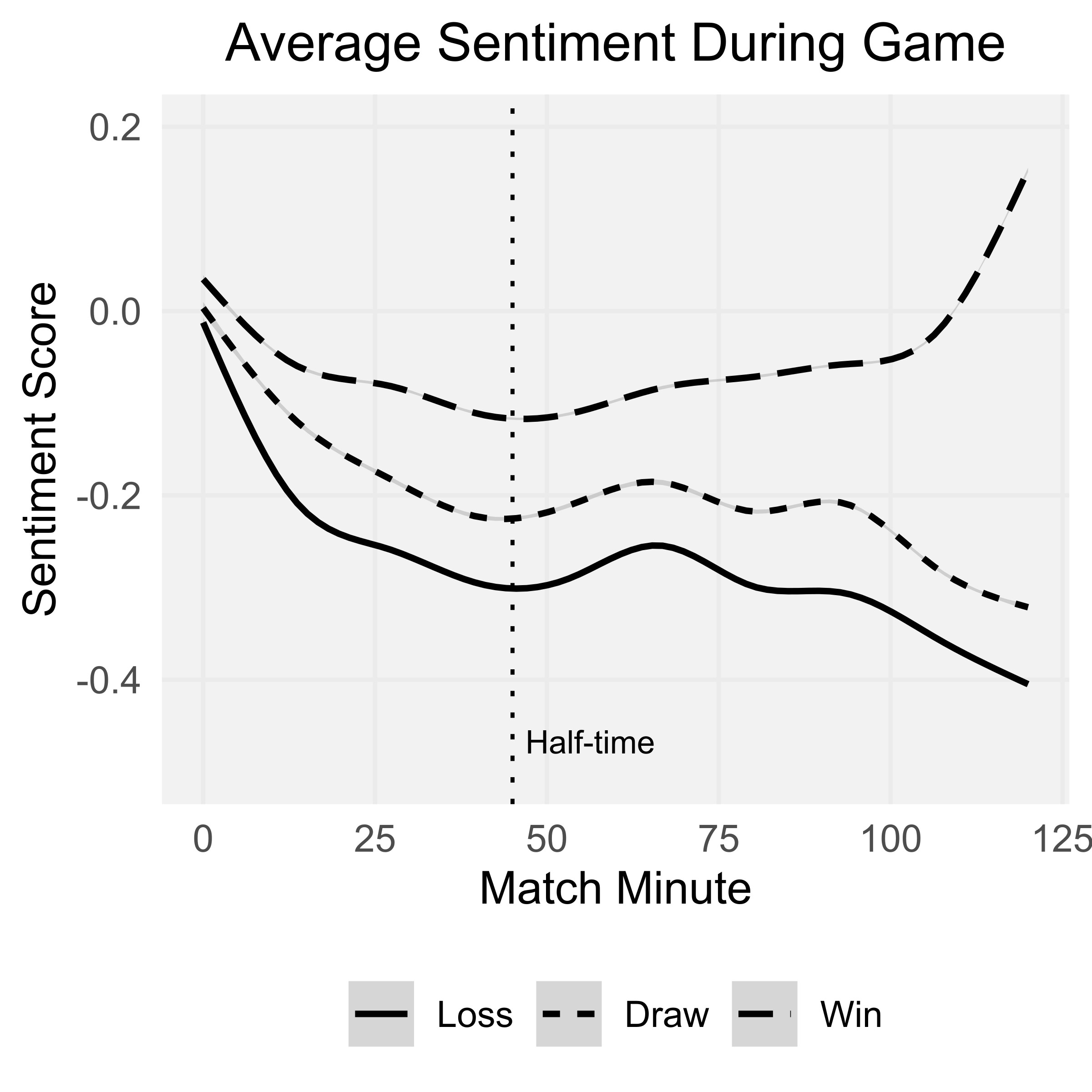}
    \caption{Aggregated sentiment change per-minute by match result (FC Corpus).}
    \label{fig:agg_by_minute}
\end{figure}

Figure \ref{fig:agg_by_minute} shows these relationships aggregated per-minute over the, roughly, 120 minutes a game takes place. Clear patterns of fan sentiment can be seen in relation to match results. Again, all three scenarios begin at similar sentiment levels. However, here we see that it universally drops at kick-off (likely representing fans' anxious dispositions during matches when final scorelines are still unknown). These patterns diverge with time. Losses and draws show progressively amplifying negativity within online communities, representing environments where negative expressions become more prevalent and/or extreme. During wins, positive sentiment remains comparatively stable and surges only towards the end, demonstrating the precarity of results in a low-scoring sport, but also how favourable outcomes rapidly transform discourse. These variations, even when aggregated, offer insights into the dynamic nature of supporter sentiment across games. Figure \ref{fig:arse_bourn} goes further by providing an overview of a single match.

\begin{figure}
    \centering
    \includegraphics[width=1\linewidth]{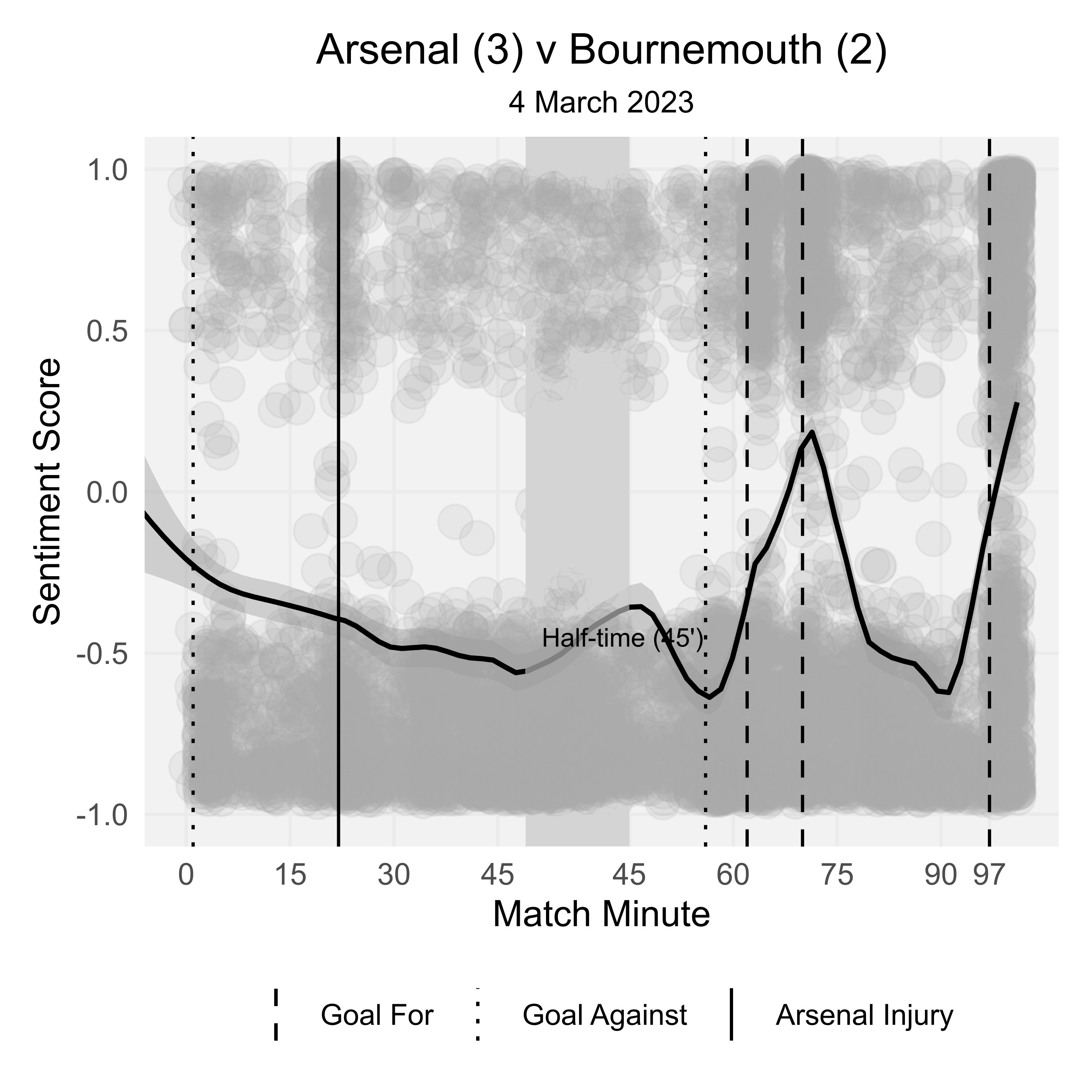}
    \caption{Post sentiment during Arsenal-Bournemouth match. Neutral posts removed to aid visualisation.}
    \label{fig:arse_bourn}
\end{figure}

Data here comes from the r/Gunners and r/ArsenalFC subreddits, and captures the dramatic March 4, 2023 Arsenal-Bournemouth match. Contextually, Arsenal needed a win to maintain their first title challenge since 2015/16, but Bournemouth scored within 9 seconds and again at 57 minutes putting Arsenal two-nil behind --- a position from which they had not comeback since 2012. However, Arsenal equalised with goals at 62 and 70 minutes before academy graduate Reiss Nelson scored a winning goal in the 97th-minute, and seconds before the whistle. 

While the figure displays mean sentiment over time, individual posts (gray dots) provide an overview of specific events. Each post is plotted by match-minute (x-axis) and sentiment score (y-axis), and the concentration of posts reveals how community sentiment clusters in terms of volume and timing in relation to live events. This is particularly visible during significant moments (marked by vertical lines), allowing one to see general ebbs and flows of sentiment, but also immediate reactions. For example, while Arsenal's first two goals trigger brief positive sentiment spikes, sentiment quickly returns to a negative baseline, suggesting negative reactions are more sustained than positive. This aligns with the asymmetric emotional response patterns seen previously (losses impacting sentiment more than wins) while also reflecting the game's high-stakes context. 

Overall this micro-level analysis reveals the dynamics of real-time emotional processing within online fan communities. However, while this is an important finding in itself, for our purposes it is further evidence of a causal relationship between football matches and online posting (although we cannot definitively establish causation from observational data alone). The consistent patterns across our aggregated and non-aggregated data, combined with previous research into fan psychology, provide compelling evidence of event-driven sentiment dynamics in online football communities. We next examine how these results influence posting habits.

\subsection{Match Results and Post Habits}

To analyse the relationship between match outcome and fan engagement, we subset data into posts made during a match (120-minutes) and up to 8-hours after, and calculated metrics to identify mean posts-per-match, the ratio between actual posts and expected posts,\footnote{Post ratio represents relative posting intensity. It is the percentage of total posts for each outcome divided by the percentage of games for that outcome. This indicates whether posting activity is higher (>1.0) or lower (<1.0) than expected based on game frequency alone.} and average sentiment (Table \ref{tab:negbin_regression_120min} and \ref{tab:negbin_regression_8hr}). The analysis reveals how match outcomes influence both posting volume and sentiment over both time periods.\footnote{Statistical significance was assessed in multiple ways. For sentiment analysis, one-way ANOVA tested overall differences in sentiment scores across match outcomes, followed by Tukey's HSD post-hoc tests to identify specific group differences. For post counts $\chi$² goodness-of-fit tests were used to determine if frequencies differed significantly from those expected. The latter results were confirmed (p < 0.001) with negative binomial regression models while controlling for season and club (subreddit) effects to account for temporal or club-specific variations that might skew aggregated results (especially as larger clubs, with more historic success, crowd out smaller clubs in the data).}

\begin{table}
  \centering
  \begin{tabular}{lccc}
   \hline
    \textbf{Result} & \makecell{\textbf{Posts per} \\ \textbf{match}} & \makecell{\textbf{Post} \\ \textbf{ratio}} & \makecell{\textbf{Average} \\ \textbf{Sentiment}} \\
    \hline
    Loss & 739.32 & 0.89*** & -0.25*** \\
    Draw & 764.37 & 0.92*** & -0.11*** \\
    Win & 934.13 & 1.12*** & 0.07*** \\
    \hline
  \end{tabular}
  \caption{Posts from all FC subreddits within 120 minutes of kick-off. $\chi$²-test for post distribution; ANOVA for sentiment differences.}
  \label{tab:negbin_regression_120min}
\end{table}
\begin{table}
  \centering
  \begin{tabular}{lccc}
   \hline
    \textbf{Result} & \makecell{\textbf{Posts per} \\ \textbf{match}} & \makecell{\textbf{Post} \\ \textbf{ratio}} & \makecell{\textbf{Average} \\ \textbf{Sentiment}} \\
    \hline
    Loss & 577.14 & 0.8*** & -0.17*** \\
    Draw & 604.96 & 0.84*** & -0.09*** \\
    Win & 904.49 & 1.26*** & 0.06*** \\
    \hline
  \end{tabular}
  \caption{Posts from all FC subreddits within 8 hours of kick-off.  $\chi$²-test for post distribution; ANOVA for sentiment.}
  \label{tab:negbin_regression_8hr}
\end{table}

Match wins generate the highest posting activity, exceeding what would be expected based on posting frequency alone (ratios of 1.12 and 1.26, representing 12\% and 26\% more posts). That is, in contrast to what we have seen up to now, victories appear to motivate fans to participate in online communities more than losses. However, the sentiment scores associated with those wins (0.07 and 0.06) are only marginally positive, while losses drive more negative sentiment (-0.25 and -0.17) despite generating (11\% and 20\%) fewer posts. 

The drop in sentiment magnitude across time-frames (during matches and up to 8-hours after) indicates that sentiment may have a decay rate. To further understand this dimension we look at these relationships over longer periods in Figure \ref{fig:sent_weeks}. The figure reports smoothed sentiment and (log) post count over time, relative to the nearest football match. This can be a match that recently happened (-1 week) or one that is upcoming (+1 week). From this perspective we can see that posting activity peaks, and sentiment drops, around matches. This pattern is somewhat surprising given that our dataset contains more wins than losses. However, it appears to be further evidence that, while wins increase the likelihood to post, losses have a stronger negative effect on sentiment. That is, on the whole, losses have a greater impact on sentiment in our dataset, and mean sentiment tends to be more positive the further a post is away from a match. Linear regression confirms this relationship: sentiment is significantly lower for posts made nearer in time to matches (p < .001).\footnote{For linear regression relative time was converted into absolute values to capture distance regardless of direction.} 

\begin{figure}
    \centering
    \includegraphics[width=1\linewidth]{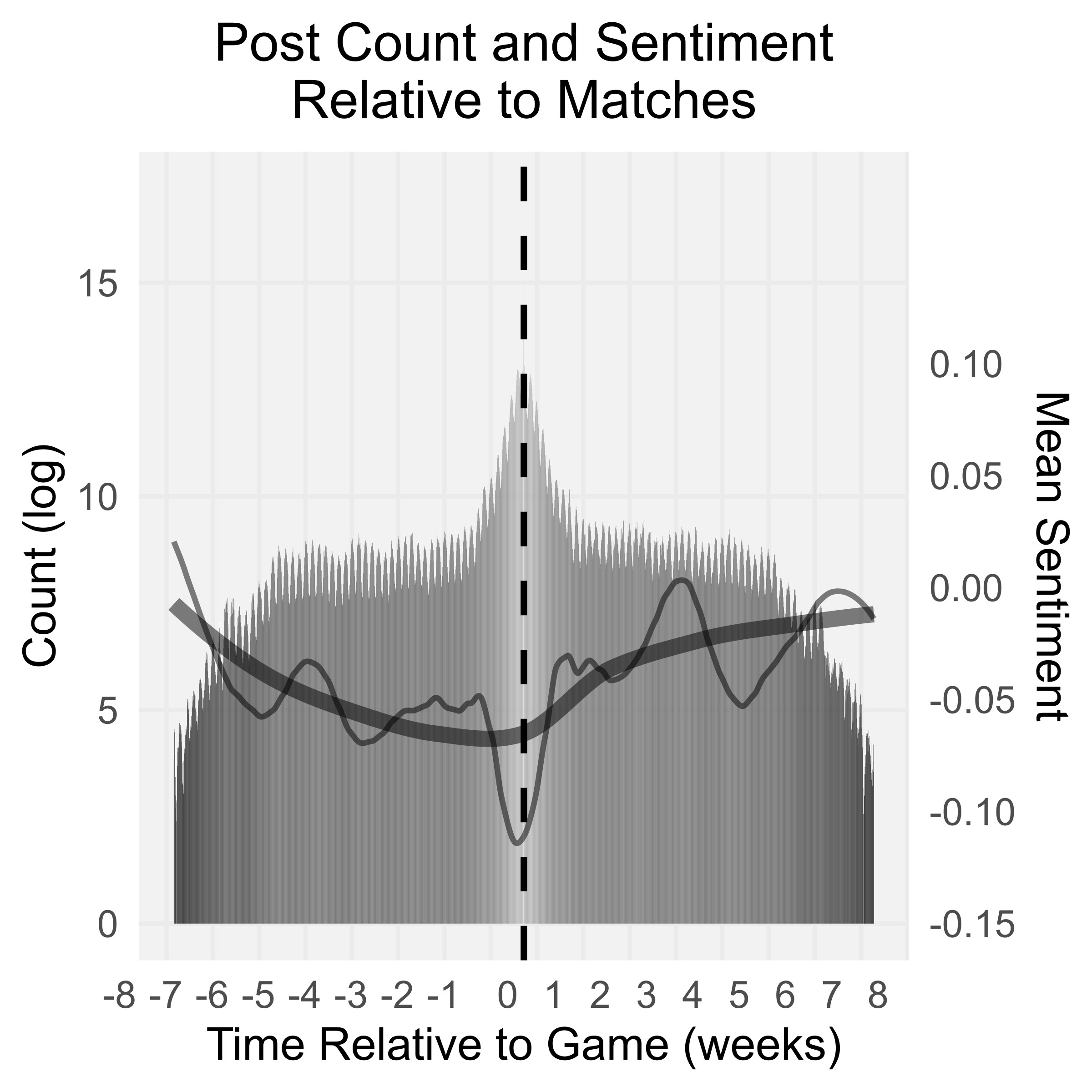}
    \caption{Post count (log) and sentiment relative to game (FC Corpus).}
    \label{fig:sent_weeks}
\end{figure}

Overall these findings reveal three key patterns complementing our match-level analyses: sentiment and posting likelihood are inversely related to match results; heightened engagement correlates with more negatively skewed sentiment; and emotional intensity diminishes over time. 

Having established a relationship between negative sentiment and problematic posting, documented the impact of matches on user posting behaviour, and quantified the magnitude of these effects, we next examine how posters interact across subreddits.

\section{Cross-Community Sentiment Spread}

Using a dataset of paired posts from club and non-club subreddits (the latter being posts made within 10-minutes of the former) we investigated how sentiment categories (negative, neutral, and positive) across post pairs are correlated (Kendall's $\tau$).\footnote{Kendall’s $\tau$ was used as it does not require data to be normally distributed. In our dataset, sentiment scores exhibit a trimodal distribution, with distinct peaks around negative, neutral, and positive values (see Figure \ref{fig:density_plots}), which violates the normality assumption required by parametric tests (such as Pearson's correlation coefficient). Additionally, as a non-parametric rank correlation measure, Kendall’s $\tau$ relies only on the ordering of the values, not their actual magnitudes (it evaluates whether sentiment scores in one context tend to align with those in another regardless of exact numerical differences, making it less susceptible to outliers or noisy data). In this context it assesses how consistently the relative ordering of one variable (e.g., sentiment in football posts) aligns with the ordering of another variable (e.g., sentiment in non-football posts). Additionally, it handles tied values better than Spearman's $\rho$, which is important given that our sentiment data falls into three discrete categories (negative, neutral, positive) and has numerous tied rankings \cite{kendall_1938}.} The analysis reveals a statistically significant --- but very weak --- correlation between post sentiment by the same users in FC and non-FC subreddits (Table \ref{tab:sentiment_with_neut}). While a finding, this in itself is not surprising. One may expect a person's emotional state to be consistent when posting across communities at similar times. However, our analysis shows the correlation doubles during matches (from $\tau$ = 0.059 to $\tau$ = 0.118). This suggests some transfer of football related emotional states to unrelated online spaces. That is, real-time football events influence sentiment in non-FC communities. 

Following existing research showing that emotionally charged content spreads online more quickly \cite{Brady_2017}, we also looked at these relationships with neutral posts made in FC subreddits removed (Table \ref{tab:sentiment_no_neut}). Here correlations during matches strengthen further ($\tau$ = 0.118 to $\tau$ = 0.146), indicating emotionally charged users are more consistent in their emotional valence across community contexts, suggesting that heightened emotional states during matches may create stronger patterns of sentiment consistency in broader online behaviour.

While the strength of the correlations remain modest in absolute terms, their relative changes and statistical significance across a large sample represent real patterns of user behaviour, and are evidence for emotional spillover.

\begin{table}
  \centering
  \resizebox{\columnwidth}{!}{
    \begin{tabular}{lrrrr}
     \hline
    \makecell{\textbf{Sentiment} \\ \textbf{Comparison}} & \makecell{\textbf{Kendall's $\tau$}} & \makecell{\textbf{n}} \\
    \hline
    \makecell{All Pairs} & 0.085*** & 575,863 \\
    \makecell{During Match} & 0.118*** & 234,024 \\
    \makecell{Outside Match} & 0.059*** & 341,839 \\
    \hline
    \end{tabular}
  }
  \caption{Statistical differences in sentiment between FC and non-FC subreddits by the same authors.}
  \label{tab:sentiment_with_neut}
\end{table}

\begin{table}
  \centering
  \resizebox{\columnwidth}{!}{
    \begin{tabular}{lrrrr}
     \hline
    \makecell{\textbf{Sentiment} \\ \textbf{Comparison}} & \makecell{\textbf{Kendall's $\tau$}} & \makecell{\textbf{n}} \\
    \hline
    \makecell{All Pairs} & 0.108*** & 354,037 \\
    \makecell{During Match} & 0.146*** & 196,957 \\
    \makecell{Outside Match} & 0.077*** & 157,080 \\
    \hline
    \end{tabular}
  }
  \caption{Statistical differences in sentiment (neutral removed) between FC and non-FC subreddits by the same authors.}
  \label{tab:sentiment_no_neut}
\end{table}

To further assess the association patterns between sentiment, a Pearson $\chi$²-test was used. Table \ref{tab:chi-2-test} reports standardised residuals, which quantify how observed sentiment pairings deviate from frequencies expected by chance (p < 0.001). The strongly positive values along the diagonal (40.00, 22.74, 44.66) demonstrate that matching sentiments occur far more frequently than expected. That is, users expressing a particular sentiment in club subreddits are more likely to express the same sentiment elsewhere. Conversely, negative residuals for mismatched pairings indicate these combinations occur less frequently than chance would predict. This provides further evidence for emotional spillover across communities.

\begin{table}
  \centering
  \resizebox{\columnwidth}{!}{
    \begin{tabular}{lrrr}
     \hline
    \makecell{\textbf{ }} & \makecell{\textbf{Negative}} & \makecell{\textbf{Neutral}} & \makecell{\textbf{Positive}} \\
    \hline
    \makecell{\textbf{Negative}} & 40.00 & -18.76 & -26.73 \\
    \makecell{\textbf{Neutral}} & -18.92 & 22.74 & -7.59 \\
    \makecell{\textbf{Positive}} & -27.49 & -5.12 & 44.66 \\
    \hline
    \end{tabular}
  }
  \caption{Pearson $\chi$² standardised residuals between sentiment pairs of FC and Non-FC subreddit posts.}
  \label{tab:chi-2-test}
\end{table}

These two tests in tandem demonstrate that there are weak but significant correlations between paired sentiments across communities and that these pairings are significantly more frequently matched than would be expected by chance, with particularly strong associations for negative-negative and positive-positive sentiment pairs. These patterns, when combined with our previous analyses, suggest that real-world events can trigger emotionally charged discourse that spreads beyond original communities. To better understand the content making this cross-community movement, we examined these posts in greater detail.

\section{Post Contents}

To understand what these paired posts linguistically represent we analysed their language features. To do this we identified posts containing profanity (taken from the LDNOOBW), violent words, intensifiers, exclamation marks, and in all-caps.\footnote{Violent words include variations of "kill", "die", "murder", "attack", "destroy", "hate", "ruin", and "merc." Intensifiers include variations of "very", "really", "so", "extremely", "absolutely", "totally", "completely", "f**king", "bloody", and "literally."} These are taken to be indicators of negative and emotionally charged discourse. Table \ref{tab:lang_tau_difference} provides an overview of correlations between categories. 

\begin{table}
  \centering
  \resizebox{\columnwidth}{!}{
    \begin{tabular}{llll}
      \hline
      \textbf{Feature} & 
      \makecell{\textbf{Outside} \\ {\textbf{Match}}} & 
      \makecell{\textbf{During} \\ {\textbf{Match}}} & 
      \makecell{\textbf{Difference} \\ ($\Delta \tau$)} \\
      \hline
      Profanity & 0.068*** & 0.124*** & 0.056 \\
      Violent & 0.033* & 0.055** & 0.022 \\
      Intensifiers & 0.072*** & 0.101*** & 0.029 \\
      Exclamations & 0.119*** & 0.154*** & 0.035 \\
      All-caps & 0.040** & 0.068*** & 0.028 \\
      \hline
    \end{tabular}
  }
  \caption{Differences in Kendall's $\tau$ between FC and non-FC subreddits for linguistic features, during-match and outside-match posts compared.}
  \label{tab:lang_tau_difference}
\end{table}
 
The table shows correlations (positive $\tau$) between a user's posts' linguistic features in different subreddits. This suggests some consistency in individual communication styles. Correlations are also stronger during matches for all features, suggesting that a user's language in FC contexts is more strongly predictive of their language in non-FC contexts while games are taking place. This strengthening of cross-subreddit linguistic patterns during matches indicates a potential causal relationship, where the heightened emotional states triggered by football events not only affect users' expressions within their particular football communities, but actively spills over to intensify their emotional communication elsewhere.

We combined ToxicityModel's probability scores with TweetNLP's sentiment scores to identify and assess posts that were tagged as toxic and highly negative. On examination many (but not all) of these would be considered highly offensive. An illustrative and anonymised set of pairs is found in Appendix \ref{sec:appendix_pairs}. These offer qualitative evidence that emotionally charged content crosses digital spaces, and that the negativity that spreads can be toxic, and thus potentially harmful. 

\section{Discussion and Conclusion}

This analysis aimed to reveal a potential causal pathway for the dissemination of negativity across digital spaces. In doing this it showed that football match outcomes appear to influence fans' emotional states, which manifest in online sentiment patterns and posting behaviours that subsequently cross community boundaries. 

The analysis provides evidence for the relationship between football match events and fluctuations in online sentiment. By examining both aggregated patterns across thousands of matches and granular minute-by-minute reactions during specific games, we observed consistent temporal alignment between on-field events and sentiment shifts in online communities, with sentiment trajectories diverging predictably based on match outcomes and in-game moments.

Further, we identified an asymmetric relationship in posting habits. In the context of football club supporter communities, this manifested as heightened negative sentiment following losses, moderate negativity after draws, and mildly positive sentiment after wins --- despite wins generating more posts on average. These findings support Kaden et al. \citeyearpar{kaden_i_2023} on the prominence of negative sentiment online, as well as Onwe's \citeyearpar{onwe2016involuntary} research showing losses have greater emotional impact on fans than wins. More broadly, they align with established psychological principles of negativity bias in emotional processing, where negative experiences tend to have more profound impacts than positive ones \cite{Roy_2001}.

Building on evidence of event-driven sentiment dynamics, we examined how these may trigger cross-community emotional spillover by analysing paired posts from the same users in club and non-club subreddits, and showed statistically significant correlations between sentiment across these community boundaries. Additionally, these correlations were stronger both during matches, and when the FC-related post was emotionally charged \cite{Brady_2017}.

Finally, following our initial analysis showing a relationship between negativity and potentially problematic language, we examined linguistic features across paired posts. We found significant correlations between specific linguistic markers indicative of negative and emotionally charged discourse. Additionally, these correlations were stronger during matches than at other times. This provides evidence that not only sentiment, but problematic language patterns, can transfer across communities with real-world events potentially triggering cascades of toxic discourse.

These findings reveal how digital spaces function not as isolated environments, but are both directly impacted by external factors (real-world events) and are themselves interconnected emotional ecosystems vulnerable to cross-domain contagion --- a phenomenon with significant implications for understanding the propagation of harmful speech online, and its existence beyond its originating contexts. 

While beyond the scope of this paper, these findings suggest several avenues for future research, including practical implications for platform moderation and design. First, the predictable nature of event-driven sentiment spillover could inform automated monitoring systems that increase vigilance or moderation thresholds within communities during high-risk periods (such as major sporting events or politically charged moments). Second, the temporal patterns identified may provide a basis for predictive models that automatically flag users exhibiting negative sentiment in external communities for enhanced monitoring. More broadly, the demonstration that sentiment and linguistic toxicity markers transfer across seemingly unrelated digital spaces suggests future work explore whether limiting cross-community mobility during emotionally charged periods could reduce harmful spillover while preserving networked discourse benefits.

In concluding, it must again be noted that football fandom has been used in this context not because it is uniquely toxic. On the contrary, analyses not included in this paper show that these subreddits are often remarkably open and welcoming spaces where personal and political topics are discussed in civil and polite manners \cite{hill2024mass}. Instead, football has been used as a case-study able to computationally measure affectional movements tied to known events. That is, football is one arena for discourse, and it is discourse itself that is central to this study. Future work will apply the methodology developed here to analyse emotional spill-over in other domains to further examine the generalisability and implications for cross-community sentiment transfer.

\section*{Limitations}

\subsection*{Contextually similar communities}

While this analysis focused on subreddits dedicated to specific football clubs --- that is, forums for supporters of a specific club rather than general football related discussions --- it should be noted that some of these football-related communities were retained in the dataset used in Section 6. These are r/soccer (dedicated to general football discussion) and r/FantasyPL (dedicated to discussion around fantasy premier league football, such as a poster's team, strategies, and results). While the statistical relationship remains with these subreddits removed (albeit with a diminished $\tau$), we reported results with them included for the following reasons.

First, while these subreddits are football-related, the communities represent neutral spaces, open to all supporters. Therefore, they are detached from the statistical relationship noted between match results and FC subreddits, and are spaces where sentiment can transfer to, despite being contextually (football) related. That is, the phenomenon we are studying is emotional spillover and if a user's club-tied sentiment spills into r/soccer, that remains of interest (especially when this spill-over represents things such as commiseration or trolling).

Second, Fantasy Football involves different emotional investments than supporting a club. In this case users themselves are the competitors. Additionally, users are known to be both pragmatic and emotionally hedge by choosing players for their fantasy teams who are signed to real-life rival teams. Therefore, there are unique emotional dynamics within this context that make it distinct from a club subreddit.

Nonetheless, contextual similarity between subreddits is an area that may obfuscate results. As context is central to any meaning --- including emotional --- it is therefore an area that should be further studied as there are likely to be implications for understanding the emotional dynamics of online communities.

\subsection*{Existing moderation}

Reddit is not static, and posts can be edited and deleted by both users and moderators. This means that our data will have missed some of the most problematic posts as users may have retroactively modified them or moderators may have removed them. In practical terms, this likely means we are under-reporting problematic posts.

\subsection*{Identifying problematic speech}

As noted in the Related Work section, there are difficulties identifying hate speech. Football discourse has its own peculiarities, with numerous words that may be more or less innocuous in a football context. To offer examples: The Liverpool player Virgil van Dijk’s name is often misspelled (sometimes purposefully and sometimes not). References to the colour of a team’s kit can set off false positives (e.g., “the whites” or “the blacks”). In our list of violent words we included "kill." However, one will often say a team is "killing the game" to mean deliberately slowing down the pace of play to maintain advantage. It would, therefore, not be surprising to find examples of these expressions more frequently during matches. Similarly, in the hate speech lexicon the word "villain" was included, which is also Aston Villa's moniker (we removed it in our case). In a similar, but more problematic, vein: many Tottenham Hotspur supporters refer to themselves as "yids" or the "yid army." While the usage may not be intended as harmful, many fans disagree (made obvious by efforts from within the club to end its usage). This adds difficulty to how we assess this word's usage. All of this is to say: while we conducted a lexical investigation into the relationship between negative sentiment and potentially problematic language, it is necessary to foreground that there are likely to be false positives within our data (despite clear correlations between negative sentiment and problematic posts).

There is also almost certainly collinearity between potentially problematic language and negative sentiment. That is, the same lexical features (profanity, hostile language, slurs) that are found in lexicons or trigger toxicity detection are also likely to influence sentiment scores. This overlap means the correlation between negative sentiment and problematic content may be partly methodological. Future work might explore how sentiment analysis approaches separate emotional valence from linguistic toxicity to better distinguish these dimensions.

\subsection*{Further opportunities to refine data}

The data refinement and matching process for this paper was extremely complicated, but can still be refined further. At the macro level we use a 120-minute time frame to assess sentiment in relation to game states, and align this with a final score. However, we have shown that at the micro level sentiment is more dynamic than this. This likely results in mismatches between sentiment and final results as users move between emotional states during a match. While this is unlikely to impact the direction of our findings, it may be causing weaker correlations and/or preventing the identification of examples. A more refined dataset in which posts are measured in relation to individual goal times and live scores would be preferable.

\section*{Acknowledgments}

The author would like to thank the anonymous peer reviewers for the WOAH at ACL 2025 Vienna; Paul Nulty for reviewing an earlier draft; Johan Ahlback for statistical sense checking; and Rafal Zaborowski for qualitative discussions on the topic.

\nocite{}

%%%%%%%% INSTRUCTIONS FOR GETTING A BBL TO WOKR:
% USE:

%\bibliography{custom}

% or whatever file. This points to custom.bib.

% Once it's rendered you can go to the logs and download the bbl file (called output.bbl). This needs to be uploaded here, and then the \bibliography{custom} needs to be commented out
% the new bbl file uploaded (and renamed)
% and the below added:

%\input{output.bbl}

\appendix
\onecolumn
\section{Appendix: Subreddits}

\label{sec:appendix_subreddits}
\begin{center}
\begin{tabular}{p{.3\linewidth}p{.3\linewidth}p{.3\linewidth}}
\hline
    \textbf{Subreddit} & \textbf{Football Club} & \textbf{Posts} \\
    \hline
    Gunners & Arsenal & 14,446,407 \\  
    reddevils & Manchester United & 13,207,365\\
    LiverpoolFC & Liverpool & 11,546,020\\
    chelseafc & Chelsea & 6,996,674\\
    coys & Tottenham Hotspur & 6,024,763\\   
    MCFC & Manchester City & 2,034,531\\
    Everton & Everton & 1,392,566  \\
    NUFC & Newcastle United & 1,260,601\\
    ManchesterUnited & Manchester United & 1,009,580\\ 
    LeedsUnited & Leeds United & 774,951\\
    Hammers & West Ham United & 770,800\\
    avfc & Aston Villa & 713,223\\
    ArsenalFC & Arsenal & 433,250\\
    SaintsFC & Southampton & 331,857\\
    lcfc & Leicester City & 244,128\\        
    nffc & Nottingham Forest & 192,217\\        
    WWFC & Wolverhampton Wanderers & 153,352\\           
    swanseacity & Swansea City & 135,281\\    
    BrightonHoveAlbion & Brighton and Hove Albion & 130,296\\
    crystalpalace & Crystal Palace & 128,474\\
    SheffieldUnited & Sheffield United & 63,606\\
    fulhamfc & Fulham & 61,451\\
    NorwichCity & Norwich City & 51,669\\      
    safc & Sunderland & 42,044\\
    Brentford & Brentford & 41,556\\ 
    Watford\_FC & Watford & 32,012 \\
    superhoops & Queens Park Rangers & 29,033\\
    WBAfootball & West Bromwich Albion & 24,126\\ 
    COYH & Luton Town & 19,590\\
    bluebirds & Cardiff City & 18,395\\    
    Burnley & Burnley & 13,843\\
    AFCBournemouth & Bournemouth & 12,660\\        
    StokeCityFC & Stoke City & 10,562\\  
    Urz & Reading & 8,749\\
    brfc & Blackburn Rovers & 6,185\\
    bcfc & Birmingham City & 5,465\\          
    HullCity & Hull City & 4,710\\     
    HuddersfieldTownFC & Huddersfield Town & 4,162\\
    latics & Wigan Athletic & 3,128\\
    bwfc & Bolton Wanderers & 2,605\\
    Boro & Middlesbrough & 2,442  \\
    \hline
\end{tabular}
\label{tab:subreddits}
\end{center}

\pagebreak

\section{Appendix: Post Pairs}
\label{sec:appendix_pairs}

\begin{center}
\fbox{
  \begin{minipage}{0.95\textwidth}
    \textbf{OFFENSIVE CONTENT WARNING}: This appendix contains some examples of hateful content. This is strictly for the purposes of enabling this research, and we have sought to minimize the number of examples where possible. Please be aware that this content could be offensive and cause you distress.
  \end{minipage}
}
\end{center}

\begin{table}[htbp] % Added [htbp] for float placement suggestion
  \centering
  \begin{tabular}{p{.45\linewidth} p{.45\linewidth}}
    \hline
    \textbf{FC Subreddit} & \textbf{Non-FC Subreddit} \\
    \hline
    "F**k our attack is completely useless" & 
    "haha you're such a f****t" (r/filmclips), "this is pure garbage" (r/photoeditbattles) \\
    "I've already told you to get f**cked you absolute c**t. Go finger your ma you pathetic f**k" & 
    "Looks like a total c*m stain that's going to produce more useless human s**t like yourself" (r/VintagePhotos), "She's ignoring you because you're a B***H" (r/maledatingadvice) \\    
    "F**k this is bulls**t" & "Eat s**t and die" (r/cambridge) \\
    "F**k off, already. Seriously, f**k off" & "Only a r****d would like this" (r/humor) \\
    "f**k. off. useless. defender." & "I hope someone violently r***s her when she
gets home (r/embarrassing) \\
    "The match fell apart when our 2-goal lead vanished because that useless goalkeeper's f**cking error." & "seriously, look at how she's dressed, total s**t" (r/elegantcelebrities) \\
    "he tried to park the bus at the home pitch. f**k off [manager]." & "eat a 40 inch c**k [player name]" (r/FootballOdds) \\
    "Pedantry can eat my d**k" & "You dumb f**king fool, you absolute s**t of a cretin, you massive f**king donkey, you bumbling idiot. F**k you." (r/HipHopFans) \\
    "happy its f**king done. I'll take a draw" & "Nope, admit you can't read you illiterate f**k" (r/socialmedia) \\
    "this is f**king our team, THIS IS OUR TEAM, WHAT A TOTAL GROUP OF F**KING MUPPETS" & "How can she be happy with herself when she's a disgusting piece of human trash?" (r/SocialMediaScreenshots) \\
    \hline
  \end{tabular}
  \caption{Example paired comments from the same author made within 10 minutes. Quotes and subreddits have been modified to avoid identification of authors.}
  \label{tab:bad_pairs}
\end{table} % Closing the table environment

\end{document}